\documentstyle[aps,prd,epsfig,preprint]{revtex}
\def\fun#1#2{\lower3.6pt\vbox{\baselineskip0pt\lineskip.9pt
\ialign{$\mathsurround=0pt#1\hfill##\hfil$\crcr#2\crcr\sim\crcr}}}
\begin{document}
\vspace{0.5in}
\title{\vskip-2.5truecm{\hfill \baselineskip 14pt 
{\hfill {{\small \hfill UT-STPD-1/99 }}}\\
{{\small \hfill THES-TP/99-02}}\vskip .1truecm} 
\vspace{1.0cm}
\vskip 0.1truecm {\bf Hierarchical Neutrinos and 
Supersymmetric Inflation }}
\vspace{1cm}
\author{{G. Lazarides}$^{(1)}$\thanks{lazaride@eng.auth.gr} 
{and N. D. Vlachos}$^{(2)}$\thanks{vlachos@physics.auth.gr}} 
\vspace{1.0cm}
\address{$^{(1)}${\it Physics Division, School of Technology, 
Aristotle University of Thessaloniki,\\ Thessaloniki GR 540 06, 
Greece.}}
\address{$^{(2)}${\it   Department of Physics, Aristotle 
University of Thessaloniki,\\Thessaloniki GR 540 06, Greece.}}
\maketitle

\vspace{2cm}

\begin{abstract}
\baselineskip 12pt

\par
A moderate extension of MSSM based on a left-right 
symmetric gauge group, within which hybrid inflation is 
`naturally' realized, is discussed. The $\mu$ problem 
is solved via a Peccei-Quinn symmetry. Light neutrinos 
acquire hierarchical masses by the seesaw mechanism. 
They are taken from the small angle MSW resolution of the 
solar neutrino puzzle and the SuperKamiokande data. 
The range of parameters consistent with maximal 
$\nu_{\mu}-\nu_{\tau}$ mixing and the gravitino 
constraint is determined. The baryon asymmetry of the 
universe is generated through a primordial leptogenesis. 
The subrange of parameters, where the baryogenesis 
constraint is also met, is specified.  The required 
values of parameters are more or less `natural'.

\end{abstract}

\thispagestyle{empty}
\newpage
\pagestyle{plain}
\setcounter{page}{1}
\def\beq{\begin{equation}}
\def\eeq{\end{equation}}
\def\beqa{\begin{eqnarray}}
\def\eeqa{\end{eqnarray}}
\def\tr{{\rm tr}}
\def\x{{\bf x}}
\def\p{{\bf p}}
\def\k{{\bf k}}
\def\z{{\bf z}}
\baselineskip 20pt

\par
It is by now clear that the minimal supersymmetric 
standard model (MSSM), being unable to provide answers to 
a number of important physical issues, must be part of a 
more fundamental theory. Some of the shortcomings of MSSM 
are the following: i) Inflation cannot be implemented in 
its context. ii) There is no understanding of how the 
supersymmetric $\mu$ term, with $\mu \sim 10^{2}-10^{3}
~{\rm{GeV}}$, arises. iii) Neutrinos remain massless and, 
consequently, there are no neutrino oscillations in 
contrast to recent experimental evidence \cite{superk}. 
iv) Last, but not least, the observed baryon asymmetry of 
the universe cannot be easily generated (through the 
electroweak sphaleron processes).

\par
Moderate extensions of MSSM provide~\cite{lyth,dss,lss,deg} 
a suitable framework within which the hybrid inflationary 
scenario \cite{hybrid} is `naturally' implemented. This
means that a) there is no need for tiny coupling constants, 
b) the superpotential can have the most general form 
allowed by symmetry, c) supersymmetry guarantees that 
radiative corrections do not invalidate inflation, but 
rather provide the necessary slope along the inflationary 
trajectory, and d) supergravity corrections can be brought 
under control.

\par
The $\mu$ problem, in these inflationary extensions of 
MSSM, could be solved \cite{dls} by coupling the inflaton 
system to the electroweak higgs superfields. In this case, 
however, the inflaton predominantly decays, after the end 
of inflation, into (s)higgses and the gravitino 
constraint \cite{gravitino} on the reheat temperature 
restricts \cite{deg,atmospheric} the relevant 
dimensionless coupling constants to `unnaturally' small 
values ($\sim 10^{-5}$). Here, we will adopt an 
alternative solution \cite{kn,rsym} to the $\mu$ 
problem that relies on coupling the electroweak higgses 
to superfields causing the breaking of the Peccei-Quinn 
\cite{pq} symmetry.

\par
Light neutrino masses can be generated either through the 
well-known seesaw mechanism after including right handed 
neutrino superfields or via coupling 
\cite{deg,rsym,triplet} the light neutrinos directly to 
$SU(2)_L$ triplet pairs of superfields of intermediate 
scale masses and `tiny' vacuum expectation values (vevs). 
The latter possibility, studied in Ref.\cite{deg}, is 
certainly more appropriate if light neutrinos are to 
provide the hot dark matter needed to explain 
\cite{structure} the large scale structure of the 
universe. Note that, for zero cosmological constant and 
almost flat spectrum of density perturbations, the hot 
dark matter density must be about 20$\%$ of the critical 
density keeping light neutrino masses in the eV range. 
Within a three neutrino scheme, compatibility with the 
atmospheric \cite{superk} and solar neutrino oscillations 
requires almost degenerate light neutrino masses which 
cannot be `naturally' obtained from the seesaw mechanism.

\par 
Measurements \cite{barfrac} of the cluster baryon 
fraction combined with the low deuterium abundance 
constraint \cite{deuterium} on the baryon asymmetry 
suggest that the matter density is around 40$\%$ of the 
critical density of the universe ($\Omega_{m}\approx 
0.4$). Also, recent observations \cite{lambda} favor 
the existence of a cosmological constant whose contribution 
to the energy density can be as large as 60$\%$ of the 
critical density ($\Omega_{\Lambda}\approx 0.6$) 
driving the total energy density close to its critical 
value as required by inflation. The assumption that 
dark matter contains only a cold component leads then 
to a `good' fit \cite{lambdafit} of the cosmic background 
radiation and both the large scale structure and age 
of the universe data. Moreover, the possibility of 
improving this fit by adding light neutrinos as hot dark 
matter appears \cite{primack} to be rather limited and, 
thus, hierarchical neutrino masses are acceptable. Note 
that neutrino masses could be hierarchical even in the 
case of zero cosmological constant provided that hot dark 
matter consists of some other particles (say axinos).

\par    
Here, we will concentrate on the case of hierarchical 
neutrino masses generated by the seesaw mechanism.
We consider a particular extension of MSSM based on a 
left-right symmetric gauge group within which hybrid 
inflation is `naturally' implemented and the $\mu$ 
problem is solved via a Peccei-Quinn symmetry. Right 
handed neutrinos are present and form $SU(2)_{R}$ 
doublets with the right handed charged leptons. The 
inflaton decays into right handed neutrino superfields, 
whose subsequent decay produces \cite{leptogenesis} a 
primordial lepton asymmetry that is later converted 
into baryon asymmetry by electroweak sphaleron effects.
  
\par
Our model is based on the left-right symmetric gauge 
group $G_{LR}=SU(3)_c\times SU(2)_L\times SU(2)_R
\times U(1)_{B-L}$~. The breaking of $SU(2)_R\times 
U(1)_{B-L}$ to $U(1)_Y$~, at a superheavy scale 
$M\sim 10^{16}~{\rm{GeV}}$, is achieved through the 
superpotential
\begin{equation}
W = \kappa S(l^c\bar l^{c}-M^2)~, 
\label{W}
\end{equation}
where $l^c$, $\bar l^{c}$ is a conjugate pair of 
$SU(2)_R$ doublet left handed superfields with $B-L$ 
charges equal to 1, -1 respectively, and $S$ is a gauge 
singlet left handed superfield. The coupling constant 
$\kappa$ and the mass parameter $M$ can be made real 
and positive by suitable phase redefinitions. The 
supersymmetric minima of the scalar potential lie on 
the D-flat direction $l^c=\bar l^{c*}$ at 
$\langle S\rangle = 0~,~ |\langle l^c \rangle| 
=|\langle \bar l^{c} \rangle| = M$.

\par
In this supersymmetric scheme, hybrid inflation 
\cite{hybrid} is automatically and `naturally' 
realized \cite{lyth,dss,lss,dls,atmospheric}. The 
scalar potential possesses a built-in inflationary 
trajectory at $l^c =\bar l^{c}=0$~, $|S|>M$ with a 
constant tree-level potential energy density 
$\kappa^{2}M^{4}$ which is responsible for the 
exponential expansion of the universe. Moreover, since 
this constant energy density breaks supersymmetry, there 
are important radiative corrections \cite{dss} which 
provide a slope along the inflationary trajectory 
necessary for driving the inflaton towards the vacua. 
At one-loop, the cosmic microwave quadrupole anisotropy 
is given \cite{dss,lss} by  
\begin{equation}
\left(\frac{\delta T}{T}\right)_{Q}\approx 4\sqrt{2}
\pi\left(\frac{N_{Q}}{45}\right)^{1/2}
\left(\frac{M}{M_{P}}\right)^{2}x_{Q}^{-1}y_{Q}^{-1}
\Lambda (x_{Q}^{2})^{-1}~,
\label{quadrupole}
\end{equation} 
where $N_Q \approx 50-60$ denotes the number of 
e-foldings experienced by our present horizon size during 
inflation, $M_{P}=1.22\times 10^{19}~{\rm{GeV}}$ is 
the Planck scale and
\begin{equation}
\Lambda (z)=
(z-1)\ln (1-z^{-1})+(z+1)\ln (1+z^{-1})~.
\label{lambda}
\end{equation} 
Also, 
\begin{equation}
y_Q^2=\int_{1}^{x_Q^{2}}\frac{dz}{z}
\Lambda(z)^{-1}~,~y_Q \geq 0~,
\label{yQ}
\end{equation}
with $x_{Q}=|S_{Q}|/M$ ($x_{Q}\geq 1$), $S_{Q}$ 
being the value of the scalar field $S$ when the scale 
which evolved to the present horizon size crossed outside 
the inflationary horizon. The superpotential parameter 
$\kappa$ can be evaluated \cite{dss,lss} from
\begin{equation} 
\kappa \approx \frac{4\sqrt{2}\pi ^{3/2}}
{\sqrt{N_{Q}}}~y_{Q}~\frac{M}{M_{P}}~\cdot
\label{kappa}
\end{equation}

The $\mu$ term can be generated \cite{dls}
by adding the superpotential coupling $S H^{2}$, where 
the electroweak higgs superfield $H=(H^{(1)}, H^{(2)})$ 
belongs to a bidoublet $(2,2)_{0}$ representation of
$SU(2)_L\times SU(2)_R\times U(1)_{B-L}$~. After 
gravity-mediated supersymmetry breaking, $S$ develops 
\cite{dls} a vev $\langle S\rangle\approx 
-m_{3/2}/\kappa$~, where $m_{3/2}\sim (0.1-1)\ 
{\rm TeV}$ is the gravitino mass, and generates a $\mu$ 
term with $\mu=\lambda \langle S\rangle\approx 
-(\lambda/\kappa)m_{3/2}$~. 

\par
This particular solution of the $\mu$ problem is 
\cite{deg,atmospheric}, however, not totally 
satisfactory since it requires the presence of 
`unnaturally' small coupling constants 
($\kappa\stackrel{_{<}}{_{\sim }}\times 10^{-5}$). 
This is due to the fact that the inflaton system 
decays predominantly into electroweak higgs superfields
via the renormalizable superpotential coupling $SH^2$. 
The gravitino constraint \cite{gravitino} on the reheat 
temperature then severely restricts the corresponding 
dimensionless coupling constant and, consequently, the 
parameter $\kappa$. Moreover, for hierarchical neutrino 
masses from the seesaw mechanism, the requirement of 
maximal $\nu_{\mu}-\nu_{\tau}$ mixing from the 
SuperKamiokande experiment \cite{superk} further reduces 
\cite{atmospheric} $\kappa$ to become of order $10^{-6}$.

\par
An alternative solution of the $\mu$ problem will be 
adopted here. It is constructed \cite{kn} by coupling 
the electroweak higgses to superfields causing the 
breaking of the Peccei-Quinn \cite{pq} symmetry 
($U(1)_{PQ}$) which solves the strong CP problem. 
For this purpose, we need two extra gauge singlet left 
handed superfields $N$ and $\bar{N}$ with $PQ$ charges 
-1 and 1. The relevant superpotential couplings are 
$N^{2} \bar{N}^2$ and $N^{2}H^{2}$. The scalar 
potential which is generated by the term 
$N^{2} \bar{N}^2$ after gravity-mediated supersymmetry 
breaking has been studied in Ref.\cite{rsym}. For a 
suitable choice of parameters, the minimum lies at 
\begin{equation}
|\langle N \rangle|~=~|\langle\bar{N}\rangle|
~\sim (m_{3/2}m_{P})^{1/2}~\sim 10^{11} {\rm{GeV}}~,
\label{vev}
\end{equation}
where $m_{P}=M_{P}/\sqrt{8\pi}$ is the `reduced' Planck
mass. This scale is identified with the symmetry breaking 
scale $f_{a}$ of $U(1)_{PQ}$~. Substitution of 
$\langle N \rangle$ in the superpotential coupling 
$N^{2}H^{2}$ generates a $\mu$ parameter of order 
$m_{3/2}$ as desired.

\par
This approach avoids the direct coupling of the inflaton 
to the electroweak higgses. Thus, the inflaton decays 
preferably to right handed neutrino superfields via 
non-renormalizable interactions, which are `naturally' 
suppressed by $m_{P}^{-1}$ (see below), rather than to 
higgses through renormalizable couplings. The gravitino 
constraint can then be satisfied for more `natural' 
values of the dimensionless coupling constants.

\par
The superpotential $W$ of the full model contains, in 
addition to the terms in Eq.(\ref{W}), the following 
couplings:
\begin{equation}
HQQ^c,~HLL^c,
~N^{2}H^{2},~N^{2} \bar{N}^2,
~\bar l^{c}\bar l^{c}L^{c}L^{c}~.
\label{couplings}
\end{equation}
Here $Q_i$ and $L_i$ are the $SU(2)_{L}$ doublet left 
handed quark and lepton superfields, whereas $Q_i^c$ and  
$L^c_i$ are the $SU(2)_{R}$ doublet antiquarks and 
antileptons ($i$=1,2,3 is the family index). The quartic 
terms in Eq.(\ref{couplings}) carry a factor $m^{-1}_P$ 
which has been left out. Also, the dimensionless coupling 
constants as well as the family indices are suppressed. 
The last (non-renormalizable) term in Eq.(\ref{couplings}) 
generates the intermediate scale Majorana masses of the 
right handed neutrinos after $SU(2)_{R}$ breaking.

\par
The continuous global symmetries of this superpotential 
are $U(1)_B$ (and, consequently, $U(1)_L$) with the 
extra superfields $S$, $l^{c}$, $\bar l^{c}$, 
$N$, $\bar{N}$ carrying zero baryon number, an 
anomalous Peccei-Quinn symmetry $U(1)_{PQ}~$, and a 
non-anomalous R-symmetry $U(1)_{R}~$. The $PQ$ and 
$R$ charges of the superfields are as follows 
($W$ carries one unit of $R$ charge):
\begin{eqnarray*}
PQ:~H(1),~Q(-1),~Q^c(0),~L(-1),~L^c(0),~S(0),
~l^{c}(0),~\bar l^{c}(0),~N(-1),~\bar{N}(1)~;
\end{eqnarray*}
\begin{equation}
R:~H(0),~Q(1/2),~Q^c(1/2),~L(1/2),~L^c(1/2),~S(1),
~l^{c}(0),~\bar l^{c}(0),~N(1/2),~\bar{N}(0)~.
\label{charges}
\end{equation}

\par
Note that $U(1)_{B}$ (and, thus, $U(1)_{L}$) 
invariance is automatically implied by $U(1)_{R}$ even 
if all possible non-renormalizable terms are included in 
the superpotential. This is due to the fact that the $R$ 
charges of the products of any three color (anti)triplets 
exceed unity and cannot be compensated since there are no 
negative $R$ charges available. 

\par
After $U(1)_{L}$ breaking by the vevs of $l^c$, 
$\bar l^c$, some lepton number violating effective 
operators will appear. In particular, the last term in 
Eq.(\ref{couplings}) will generate the desirable 
intermediate scale masses for the right handed neutrinos. 
However, undesirable mixing of the higgs $H^{(2)}$ with 
$L$ 's will also emerge from the allowed superpotential 
couplings $N\bar{N}LHl^c$ after the breaking of 
$U(1)_{PQ}$ by the vevs of $N$, $\bar{N}$. Also, the 
$L$ 's will mix with $\bar l^c$ via the allowed 
couplings $N\bar{N}L^c\bar l^c$. To avoid such 
complications, we impose an extra discrete $Z_2$ 
symmetry (`lepton parity') under which $L$, $L^c$ 
change sign.

\par 
The only superpotential terms which are permitted by the 
global symmetries $U(1)_{R}$~, $U(1)_{PQ}$ and `lepton 
parity' are the ones in Eqs.(\ref{W}) and 
(\ref{couplings}) as well as $LLl^cl^c\bar{N}^2 l^c
\bar l^c$ and $LLl^cl^cHH$ modulo arbitrary 
multiplications by non-negative powers of the combination 
$l^c\bar l^c$. The vevs $\langle N \rangle,
~\langle\bar{N}\rangle$ together break $U(1)_{PQ} 
\times U(1)_{R}$ completely. Note that $U(1)_{L}$ is 
broken completely together with the gauge $U(1)_{B-L}$ by 
the superheavy vevs of $l^c$, $\bar l^c$. Thus, only 
$U(1)_{B}$ and `lepton parity' remain exact. 

\par
As indicated above, after lepton number violation, 
the last term in Eq.(\ref{couplings}) generates 
intermediate scale masses for the right handed neutrino 
superfields $\nu^{c}_i$ ($i$=1,2,3). The dimensionless 
coupling constant matrix of this term can be made 
diagonal with positive entries $\gamma_i$ ($i$=1,2,3) 
by a rotation on $\nu^{c}_i$ 's. The right handed 
neutrino mass eigenvalues are then 
$M_i=2\gamma_i M^{2}/m_{P}$ 
(with $\langle l^{c}\rangle$, $\langle \bar l^{c} 
\rangle$ taken positive by a $B-L$ transformation). 

\par
The light neutrino masses are generated via the seesaw 
mechanism in our scheme and, therefore, cannot be 
`naturally' degenerate. We will, thus, assume hierarchical 
light neutrino masses. Analysis \cite{giunti} of the 
CHOOZ experiment \cite{chooz} shows that the oscillations 
of solar and atmospheric neutrinos decouple. This fact 
allows us to concentrate on the two heaviest families 
ignoring the first one. We will denote the two positive 
eigenvalues of the light neutrino mass matrix by $m_{2}$ 
(=$m_{\nu _{\mu }}$), $m_{3}$ (=$m_{\nu _{\tau }}$). 
We take $m_{\nu_{\mu}}\approx 2.6\times 
10^{-3}~\rm{eV}$ which is the central value of the 
$\mu$-neutrino mass coming from the small angle MSW 
resolution of the solar neutrino problem \cite{smirnov}. 
The $\tau$-neutrino mass is taken to be 
$m_{\nu _{\tau }}\approx 7\times 10^{-2}~\rm{eV}$ 
which is the central value implied by SuperKamiokande 
\cite{superk}. 

\par
The determinant and the trace invariance of the light 
neutrino mass matrix imply\cite{neu} two constraints 
on the (asymptotic) parameters which take the form: 
\begin{equation}
m_{2}m_{3}\ =\ \frac{\left( m_{2}^{D}m_{3}^{D}
\right) ^{2}}{M_{2}\ M_{3}}~,
\label{determinant}
\end{equation}
\begin{eqnarray*}
m_{2}\,^{2}+m_{3}\,^{2}\ =\frac{\left( m_{2}^{D}\,\,
^{2}{\rm c}^{2}+m_{3}^{D}\,^{2}{\rm s}^{2}\right) 
^{2}}{M_{2}\,^{2}}+
\end{eqnarray*}
\begin{equation}
\ \frac{\left( m_{3}^{D}\,^{2}{\rm c}^{2}+m_{2}^{D}\,
^{2}{\rm s}^{2}\right)^{2}}{M_{3}\,^{2}}+\ 
\frac{2(m_{3}^{D}\,^{2}-m_{2}^{D}\,^{2})^{2}
{\rm c}^{2}{\rm s}^{2}\,{\cos 2\delta }}
{M_{2}\,M_{3}}~\cdot
\label{trace} 
\end{equation}
Here, ${\rm c}=\cos \theta ,\ {\rm s}=\sin \theta $, 
and $\theta$ and $\delta$ are the rotation angle and 
phase which diagonalize the Majorana mass matrix of the 
right handed neutrinos in the basis where the `Dirac' 
mass matrix of the neutrinos is diagonal with (positive) 
eigenvalues $m_{2}^{D}$, $m_{3}^{D}$ 
($m_{2}^{D}\leq m_{3}^{D}$).

\par
The $\mu-\tau$ mixing angle $\theta _{\mu\tau}$ 
lies \cite{neu} in the range
\begin{equation}
|\,\varphi -\theta ^{D}|\leq \theta _{\mu\tau}\leq
\varphi +\theta ^{D},\ {\rm {for}\ \varphi +
\theta }^{D}\leq \ \pi /2~,
\label{mixing}
\end{equation}
where $\varphi$ is the rotation angle which diagonalizes 
the light neutrino mass matrix in the basis where the 
`Dirac' mass matrix is diagonal and $\theta ^{D}$ is the 
`Dirac' mixing angle (i.e., the `unphysical' mixing angle 
with zero Majorana masses for the right handed neutrinos).

\par
We now turn to the discussion of the inflaton decay. Here 
the inflaton consists \cite{lss} of the two complex scalar 
fields $S$ and $\theta=(\delta\phi+\delta\bar{\phi})
/\sqrt{2}$ ($\delta\phi=\phi-M$, $\delta\bar{\phi}=
\bar{\phi}-M$ with $\phi$, $\bar{\phi}$ being the 
neutral components of $l^{c}$, $\bar{l^{c}}$) with a 
common mass $m_{infl}=\sqrt{2}\kappa M$. The scalar $S$ 
($\theta$) can decay into a pair of bosonic (fermionic) 
$\nu^{c}_{i}$ 's via the last coupling in 
Eq.(\ref{couplings}) and the coupling 
$\kappa Sl^{c}\bar{l^{c}}$. The decay width is 
the same for both scalars and equals
\begin{equation} 
\Gamma=\frac{1}{8\pi}~\left(\frac{M_{i}}
{M}\right)^{2}m_{infl}~.
\label{width}
\end{equation}
Of course, decay of the inflaton into $\nu^{c}_i$ is only 
possible if $M_i < m_{infl}/2$. The reheat 
temperature is then given \cite{lss} by
\begin{equation}
T_r\ \approx \frac{1}{7} 
\left( \Gamma M_P\right)^{1/2}~, 
\label{reheat}
\end{equation}
for MSSM spectrum, and the gravitino constraint 
\cite{gravitino} 
($T_r\stackrel{_{<}}{_{\sim }}10^9$ GeV) implies 
strong bounds on the $M_i$ 's which satisfy the inequality 
$M_i < m_{infl}/2$. Consequently, the corresponding 
dimensionless coupling constants, $\gamma_{i}$~, are 
restricted to be quite small.

\par
\begin{figure}
\epsfig{figure=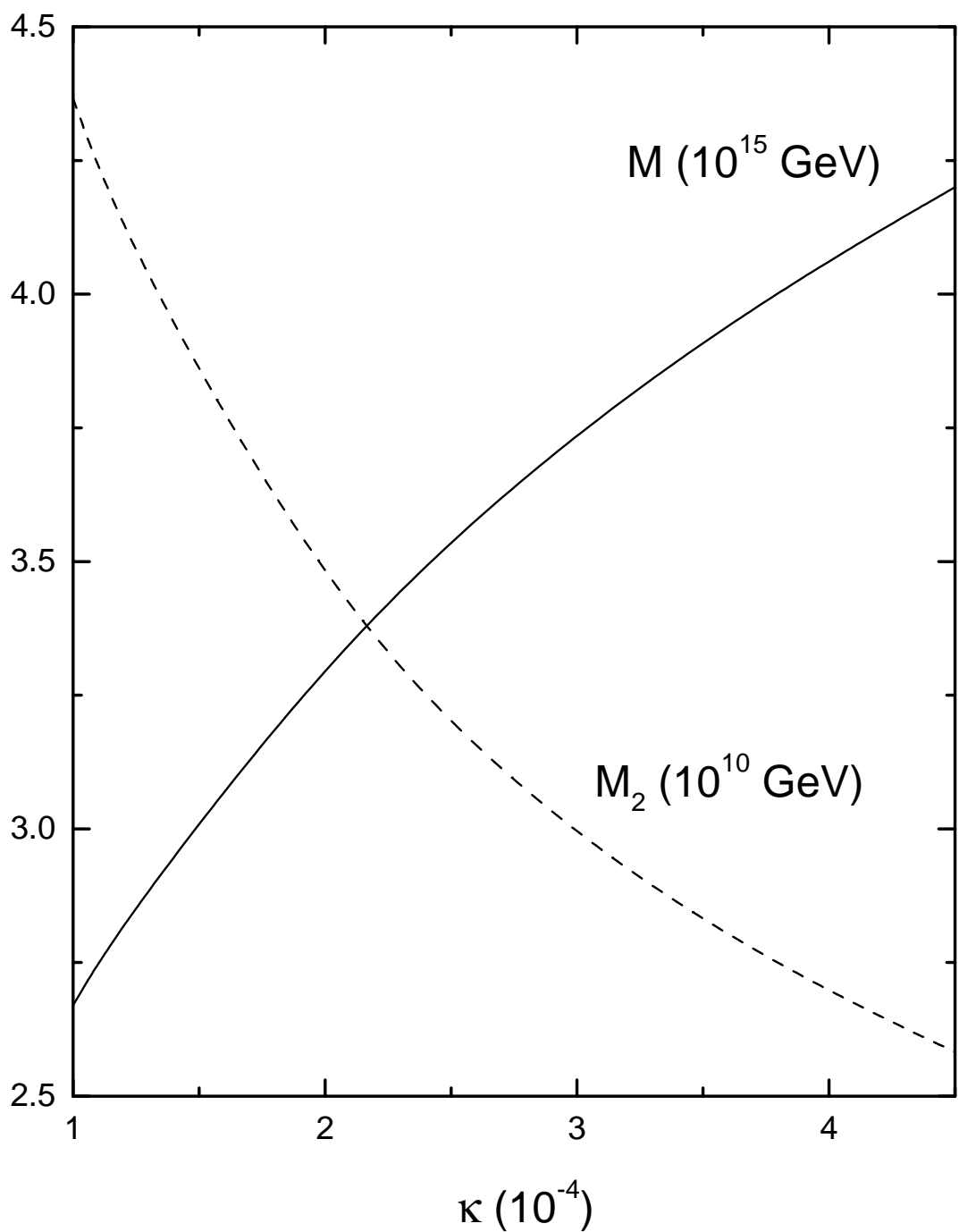,height=7.5in,angle=0}
\medskip
\caption{The mass scale $M$ (solid line) and the Majorana 
mass of the second heaviest right handed neutrino $M_{2}$ 
(dashed line) as functions of $\kappa$.
\label{M}}
\end{figure}

To minimize the number of small couplings, we take 
$M_2 < m_{infl}/2\leq M_3$ so that the inflaton 
decays into only one (the second heaviest) right handed 
neutrino with mass $M_2$. Moreover, we take 
$\gamma_{3}=1$, which gives $M_{3}=2M^{2}/m_{P}$. 
Using Eq.(\ref{kappa}), the requirement 
$m_{infl}/2\leq M_3$ becomes 
$y_{Q}\leq \sqrt{2N_{Q}}/\pi \approx 3.34$,
for $N_{Q}=55$, and Eq.(\ref{yQ}) gives 
$x_{Q}\stackrel{_{<}}{_{\sim }} 3.5$. 
Eqs.(\ref{quadrupole}) and (\ref{kappa}) with 
$(\delta T/T)_{Q}\approx 6.6\times 10^{-6}$ from 
COBE \cite{cobe} are then used to calculate $M$ and 
$\kappa$ for each value of $x_{Q}$ in this range. 
Eliminating $x_{Q}$, we obtain $M$ as a function of 
$\kappa$ depicted in Fig.\ref{M}. The inflaton mass 
$m_{infl}$ and the heaviest right handed neutrino mass 
$M_{3}$ are readily evaluated. The mass of the second 
heaviest right handed neutrino $M_{2}$ is restricted by 
the gravitino constraint \cite{gravitino} on $T_{r}$. 
We take it to be equal to its maximal allowed value in 
order to maximize $\gamma_{2}$. The value of $M_{2}$ 
is also depicted in Fig.\ref{M}.

\par
In our model, baryon number is conserved up to `tiny' 
non-perturbative $SU(2)_L$ instanton effects. So the 
observed baryon asymmetry of the universe 
can only be produced by first generating a primordial 
lepton asymmetry \cite{leptogenesis} which is then 
partially converted into baryon asymmetry by 
non-perturbative electroweak sphaleron effects. The 
lepton asymmetry is produced through the decay of the 
superfield $\nu^{c}_{2}$ which emerges as decay product 
of the inflaton. This mechanism for leptogenesis has been 
discussed in Ref.\cite{leptogenesis}. The $\nu^{c}_{2}$ 
superfield decays into electroweak higgs and (anti) lepton 
superfields. The relevant one-loop diagrams are both of the 
vertex and self-energy type \cite{covi} with an exchange 
of $\nu^{c}_{3}$. The resulting lepton asymmetry 
is \cite{neu}
\begin{equation}
\frac{n_{L}}{s}\approx 1.33~\frac{9T_{r}}
{16\pi m_{infl}}~\frac{M_2}{M_3}
~\frac{{\rm c}^{2}{\rm s}^{2}\ 
\sin 2\delta \ 
(m_{3}^{D}\,^{2}-m_{2}^{D}\,^{2})^{2}}
{|\langle H^{(1)}\rangle|^{2}~(m_{3}^{D}\,^{2}\ 
{\rm s}^{2}\ +
\ m_{2}^{D}\,^{2}{\rm \ c^{2}})}~,
\label{leptonasym}
\end{equation}
where $|\langle H^{(1)}\rangle|\approx 174~\rm{GeV}$.
Note that this formula holds \cite{pilaftsis} provided 
that $M_{2}\ll M_{3}$ and the decay width of 
$\nu^{c}_{3}$ is much smaller than 
$(M_{3}^{2}-M_{2}^{2})/M_{2}$, 
and both conditions are well satisfied here. For MSSM 
spectrum, the observed baryon asymmetry is 
given \cite{ibanez} by $n_{B}/s=-(28/79)(n_{L}/s)$.
It is important to ensure that the primordial lepton 
asymmetry is not erased by lepton number violating 
$2\rightarrow 2$ scattering processes at all 
temperatures between $T_r$ and 100 GeV. This gives 
\cite{ibanez} $m_{\nu_{\tau}}\stackrel{_<}{_\sim} 
10~{\rm{eV}}$ which is readily satisfied.

\par
\begin{figure}
\epsfig{figure=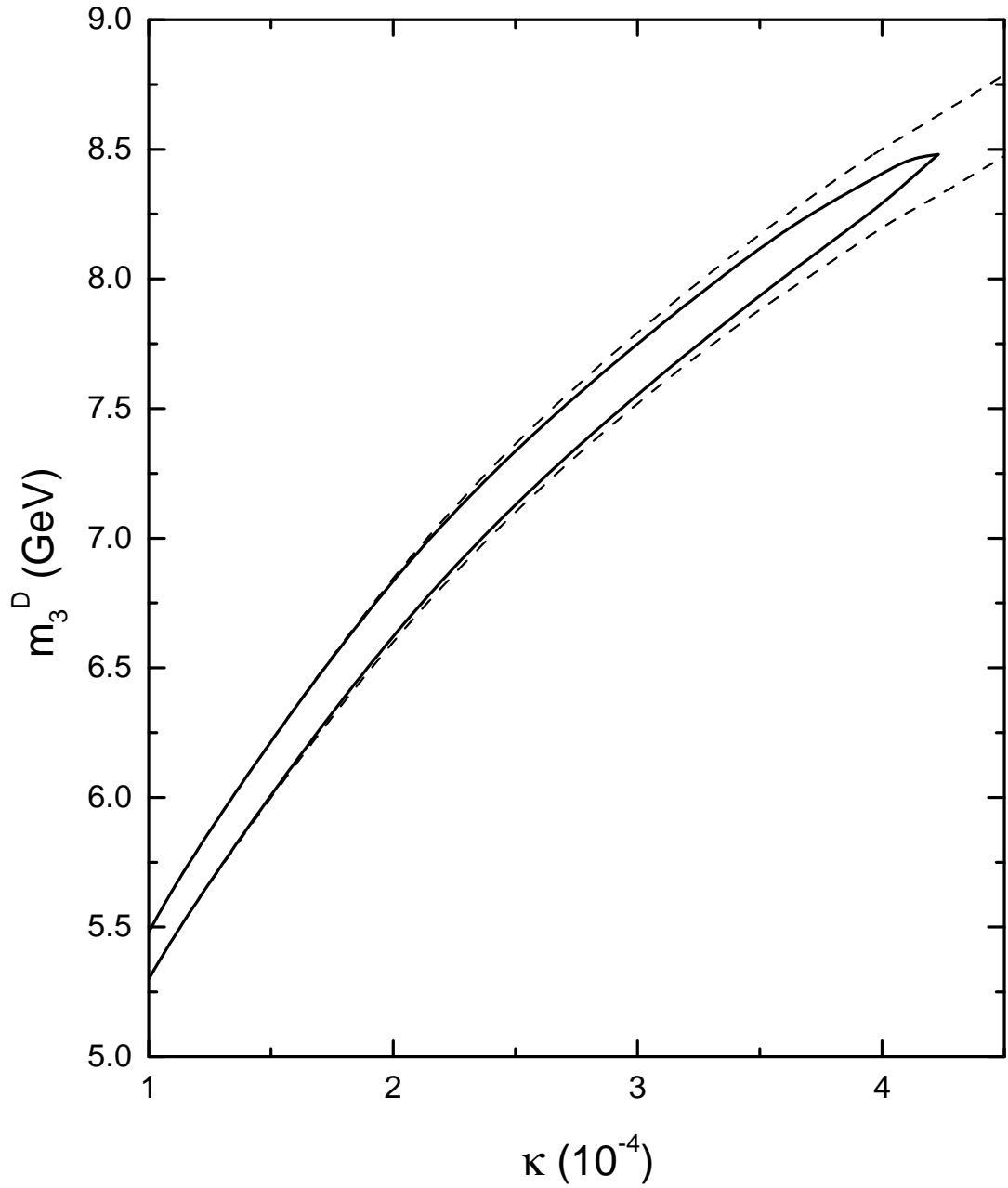,height=7in,angle=0}
\medskip
\caption{The area (bounded by the dashed lines) on the 
$\kappa-m^{D}_{3}$ plane consistent with maximal 
$\nu_{\mu}-\nu_{\tau}$ mixing and the gravitino 
constraint. Along the thick solid line the low deuterium 
abundance constraint on the baryon asymmetry of the 
universe is also satisfied.
\label{mD3}}
\end{figure}

For definiteness, we assume that the $\nu_{\mu}-
\nu_{\tau}$ mixing is about maximal 
($\theta _{\mu\tau}\approx\pi/4$) in accordance to 
the recent SuperKamiokande data \cite{superk}. We will 
also make the plausible assumption that the `Dirac' mixing 
angle $\theta^{D}$ is negligible 
($\theta^{D}\approx 0$). Under these circumstances, the 
rotation angle  $\varphi\approx\pi/4$. Using the 
`determinant' and `trace' constraints in 
Eqs.(\ref{determinant}) and (\ref{trace}) and 
diagonalizing the light neutrino mass matrix, we can 
determine the range of $m^{D}_{3}$ which allows maximal 
$\nu_{\mu}-\nu_{\tau}$ mixing for each value of $\kappa$.
These ranges are depicted in Fig.\ref{mD3} for all 
relevant values of $\kappa$ and constitute the area in 
the $\kappa-m^{D}_{3}$ plane consistent with maximal 
mixing. For each allowed pair $\kappa$, $m^{D}_{3}$, 
the value of the phase $\delta$ leading to maximal 
mixing can be determined from the `trace' condition. 
The corresponding lepton asymmetry is then found from
Eq.(\ref{leptonasym}). The line consistent with the
low deuterium abundance constraint \cite{deuterium} on 
the baryon asymmetry of the universe ($\Omega_{B}h^2
\approx 0.019$) is also depicted in Fig.\ref{mD3}. We
see that the required values of $\kappa$ ($\stackrel{_<}
{_\sim}4.2\times 10^{-4}$), although somewhat small, are
much more `natural' than the ones encountered in previous 
models \cite{atmospheric} that solved the $\mu$ problem 
and achieved maximal $\nu_{\mu}-\nu_{\tau}$ mixing. 
For these values of $\kappa$, the parameter $\gamma_{2}
\stackrel{_>}{_\sim}1.9\times 10^{-3}$, which is quite 
satisfactory.

\par
We have presented a moderate extension of MSSM based on 
a left-right symmetric gauge group. Hybrid inflation is 
`naturally' realized and the $\mu$ problem is solved 
via a Peccei-Quinn symmetry. Light neutrinos 
acquire hierarchical masses by the seesaw mechanism, 
which are taken to be consistent with the small angle 
MSW resolution of the solar neutrino problem and the
SuperKamiokande data. Under these conditions, we 
determine the range of $\kappa$ and the (asymptotic) 
`Dirac' neutrino masses consistent with 
maximal $\nu_{\mu}-\nu_{\tau}$ mixing and the 
gravitino constraint. The baryon asymmetry of the 
universe is generated through a primordial leptogenesis.
We specify the subrange of parameters where the 
baryogenesis constraint is also met. The required 
values of the relevant coupling constants turn out to be 
more or less `natural'.

\def\ijmp#1#2#3{{ Int. Jour. Mod. Phys. }{\bf #1~}(19#2)~#3}
\def\pl#1#2#3{{ Phys. Lett. }{\bf B#1~}(19#2)~#3}
\def\zp#1#2#3{{ Z. Phys. }{\bf C#1~}(19#2)~#3}
\def\prl#1#2#3{{ Phys. Rev. Lett. }{\bf #1~}(19#2)~#3}
\def\rmp#1#2#3{{ Rev. Mod. Phys. }{\bf #1~}(19#2)~#3}
\def\prep#1#2#3{{ Phys. Rep. }{\bf #1~}(19#2)~#3}
\def\pr#1#2#3{{ Phys. Rev. }{\bf D#1~}(19#2)~#3}
\def\np#1#2#3{{ Nucl. Phys. }{\bf B#1~}(19#2)~#3}
\def\mpl#1#2#3{{ Mod. Phys. Lett. }{\bf #1~}(19#2)~#3}
\def\arnps#1#2#3{{ Annu. Rev. Nucl. Part. Sci. }{\bf
#1~}(19#2)~#3}
\def\sjnp#1#2#3{{ Sov. J. Nucl. Phys. }{\bf #1~}(19#2)~#3}
\def\jetp#1#2#3{{ JETP Lett. }{\bf #1~}(19#2)~#3}
\def\app#1#2#3{{ Acta Phys. Polon. }{\bf #1~}(19#2)~#3}
\def\rnc#1#2#3{{ Riv. Nuovo Cim. }{\bf #1~}(19#2)~#3}
\def\ap#1#2#3{{ Ann. Phys. }{\bf #1~}(19#2)~#3}
\def\ptp#1#2#3{{ Prog. Theor. Phys. }{\bf #1~}(19#2)~#3}
\def\plb#1#2#3{{ Phys. Lett. }{\bf#1B~}(19#2)~#3}
\def\apjl#1#2#3{{Astrophys. J. Lett. }{\bf #1~}(19#2)~#3}
\def\n#1#2#3{{ Nature }{\bf #1~}(19#2)~#3}
\def\apj#1#2#3{{Astrophys. Journal }{\bf #1~}(19#2)~#3}
\def\anj#1#2#3{{Astron. J. }{\bf #1~}(19#2)~#3}
\def\mnras#1#2#3{{MNRAS }{\bf #1~}(19#2)~#3}
\def\grg#1#2#3{{Gen. Rel. Grav. }{\bf #1~}(19#2)~#3}
\def\ibid#1#2#3{{ibid. }{\bf #1~}(19#2)~#3}

\newpage

\pagestyle{empty}

\end{document}